\documentclass[aps,prl,floats,floatfix,twocolumn]{revtex4}

\usepackage{ifthen}
\usepackage{ifpdf}
\usepackage{color}

\ifpdf
\usepackage{graphicx}
\usepackage{epstopdf}
\else
\usepackage{graphicx}
\usepackage{epsfig}
\fi

\usepackage{latexsym}
\usepackage{amsmath}
\usepackage{amssymb}
\usepackage{bm}
\usepackage{wasysym}

\begin{document}
\title{Self-trapping of excitations: Two-dimensional quasiparticle solitons in an extended Bose-Hubbard dimer array}
\author{Amit Dey}
\author{Amichay Vardi} 
\affiliation{Department of Chemistry,
Ben-Gurion University of the Negev, Beer-Sheva 84105, Israel}
\date{\today}
\begin{abstract}
Considering a two-dimensional Bose-Hubbard spinor lattice with weak nearest neighbour interactions and no particle transfer between sites, we theoretically study the transport of energy from one initially excited dimer, to the rest of the lattice.  Beyond a critical interaction strength, low energy on-site excitations are quickly dispersed throughout the array, while  stronger excitations are self trapped, resulting in localized energy breathers and solitons. These structures are quasiparticle analogues to the discrete 2D solitons in photonic lattices. Full many-body simulations additionally demonstrate the localization of one-particle entropy.
\end{abstract}
\maketitle

Depending upon the competing interaction scales, the dynamics of bosonic Josephson junctions (BJJ) composed of two coupled Bose-Einstein condensates (BECs), give rise to intriguing phenomena such as Josephson oscillation \cite{milburn,levy07} and macroscopic quantum self-trapping \cite{raghavan,smerzi,albeiz,abbarchi}. The first phenomenon is a signature of macroscopic phase coherence already realized in superconductor Josephson junctions and in superfluid helium \cite{simmonds};  whereas, the second emerges as a consequence 
of interactions among the condensate particles.  Weak nonlinear inter-particle interaction results in anharmonic Josephson oscillations between the two condensates. When the interaction is sufficiently strong, a new dynamical self-trapping regime appears \cite{albeiz,raghavan}. Thus, while low population imbalance preparation still gives full population oscillations between the condensates, larger initial population differences remain self trapped. 

This BJJ dynamics is captured well by a tight-binding two-mode Bose-Hubbard Hamiltonian (BHH) \cite{gati,milburn,vardi,bloch}.  The BHH model has one characteristic interaction parameter $u\equiv UN/K$ with $U$,$N$, and $K$  being the interaction strength, the total number of atoms, and the hopping frequency, respectively. In the classical, mean-field limit, when $u>1$, the $\{n,\varphi\}$ phase space, where $n$ and $\varphi$ are the population imbalance and relative phase between the modes, is split by a separatrix trajectory into a Josephson oscillation region and a self-trapping region \cite{milburn,vardi,zibold10,gerving12}. If $n(t=0)$ is above the separatrix, one obtains self-trapping.  Similar self-trapped solutions can be found for attractive interaction ($u<-1$) with the same population imbalance and an opposite phase.

Going beyond a single BJJ, a system of two coupled Bose Hubbard dimers was considered in Refs. \cite{anglin1,strzys12a,strzys12b,chianca11,khripkov13,khripkov14}. The low energy dynamics of this configuration was shown to involve two Josephson modes in which both particles and elementary Bogoliubov excitations (termed 'josons' in \cite{anglin1}) are transferred between the dimers. The nonlinear repulsive interaction between the atoms was shown to result in an attraction between these quasiparticles. In a two-dimensional array of weakly coupled BH dimers, the joson 
attraction leads to their spontaneous accumulation which, on identification of excitations as `heat', emulates negative specific heat in cold atom systems \cite{anglin3}. 
 
Since the possibility to transfer excitation energy as well as particles opens the way for a second Josephson oscillation mode, it is natural to ask whether excitation energy can also be self trapped.  Here, we explore this possibility, using an extended Bose-Hubbard dimer array model. We consider $N$ particles in a two-dimensional $M\times M$ array of spinor BECs, \cite{cataliotti01} where each site has two internal degrees of freedom. To focus on the dynamics of excitations rather than on the transport of particles, there is no particle transfer between the dimers, yet energy transfer is allowed via weak long range interaction between particles in adjacent sites. This extended Bose-Hubbard model corresponds to the tight-binding description of BECs of particles with permanent magnetic or electric dipole moments, confined by periodic potentials \cite{Dutta15}. 

Launching the system with one excited site, whereas all other sites are in their respective ground states, we study the transport of the excitation energy within the array. For sufficiently strong interaction, excitations beyond a critical energy are self-trapped, resulting in energy breathers and solitons, while weaker on-site excitations are rapidly dispersed throughout the array. These excitation solitons are analogous to the two-dimensional discrete solitons obtained in photonic lattices \cite{fleischer03} and suggested for BECs in optical lattices \cite{gligoric10}. However, in our case quasiparticles (namely josephson excitations) are localized instead of particles. This leads to the localization of one-particle entropy as well as of energy, distinguishing the new structures as `heat solitons'. 
 
 The extended BHH for our system reads,
\begin{eqnarray}
 H&=&H_0+H_c,
 \label{ham}
\end{eqnarray}
where the uncoupled dimer Hamiltonian $H_0$ and the coupling Hamiltonian $H_c$ are given by
\begin{equation}
 H_0=\sum_{i}\Big[-\frac{K}{2}(\hat{a}^{\dagger}_{i}\hat{b}_{i}+\hat{b}^{\dagger}_i\hat{a}_i)+\frac{U}{2}(\hat{n}^2_{a,i}+\hat{n}^2_{b,i})\Big],
 \label{ham1}
\end{equation}
and
\begin{eqnarray}
 H_c&=&\frac{\mathcal{U}_{aa}}{2}\sum_{<i,j>}(\hat{n}_{a,i}\hat{n}_{a,j}+\hat{n}_{b,i}\hat{n}_{b,j})\nonumber \\
 ~&~&+\frac{\mathcal{U}_{ab}}{2}\sum_{<i,j>}(\hat{n}_{a,i}\hat{n}_{b,j}+\hat{n}_{b,i}\hat{n}_{a,j}),
 \label{int}
\end{eqnarray}
respectively. In the above, $\hat{a}_i$ and $\hat{b}_i$ are the bosonic destruction operators for the two dimer modes 
at site $i\equiv\{i_1,i_2\}$ and the respective mode populations are $\hat{n}_{a,i}=\hat{a}^{\dagger}_i\hat{a}_i$, $\hat{n}_{b,i}=\hat{b}^{\dagger}_i\hat{b}_i$. The intra-dimer coupling strength between the two modes and the on-site inter-particle interaction strength are given by $K$ and $U$, whereas $\mathcal{U}_{aa}$ and $\mathcal{U}_{ab}$ 
are same-mode and opposite-mode nearest-neighbor interaction strengths, respectively. 
We choose parameter values $K,U>>\mathcal{U}_{aa}, \mathcal{U}_{ab}$ to have time-scale separation between 
the fast intra-dimer and slow inter-dimer dynamics.  Since no inter-site particle transfer is allowed,  
particle numbers $n_i=n_{a,i}+n_{b,i}$ are separately conserved at each site.  Below we assume a uniform particle distribution $n_i=n$.

Equation (\ref{ham}) can be mapped into a Heisenberg-like spin Hamiltonian by defining the SU(2) local spin operators $ \hat{L}_{+,i}=\hat{a}^{\dagger}_{i}\hat{b}_{i}$, $\hat{L}_{z,i}=(\hat{n}_{a,i}-\hat{n}_{b,i})/2$ \cite{milburn,vardi}. Neglecting insignificant c-numbers $H_0$ and $H_c$ become,
\begin{eqnarray}
 H_0&=&\sum_{i}(-K\hat{L}_{x,i}+U\hat{L}^2_{z,i}) \\
 H_c&=&\mathcal{U}\sum_{<ij>} \hat{L}_{z,i}\hat{L}_{z,j},
\end{eqnarray}
where $\mathcal{U}=\mathcal{U}_{aa}-\mathcal{U}_{ab}$. Since $[{\hat L}^2,H]=0$, the local spin magnitude is conserved at $\ell=\frac{n}{2}$.  Normalizing the spin operators as $\hat {\bf s} \equiv {\hat L}/\ell$, the Heisenberg equation of motion are,
\begin{eqnarray}
\frac{d{{\hat s}_{x,i}}}{d\tau}&=&-{\rm u}{\hat s}_{z,i}{\hat s}_{y,i}-{\rm v}\sum_{<j>} {\hat s}_{z,j}{\hat s}_{y,i}, 
 \label{bloch_eqn1}\\
 \frac{d{\hat s}_{y,i}}{d\tau}&=&{\hat s}_{z,i}+{\rm u}{\hat s}_{z,i}{\hat s}_{x,i}+{\rm v}\sum_{<j>} {\hat s}_{z,j}{\hat s}_{x,i},
 \label{bloch_eqn2}\\
 \frac{d{\hat s}_{z,i}}{d\tau}&=&-{\hat s}_{y,i},
\label{bloch_eqn3}
 \end{eqnarray}
 with dimensionless time $\tau=Kt$ and the dimensionless interaction parameters ${\rm u}\equiv\frac{Un}{K}$ and 
 ${\rm v}\equiv \frac{\mathcal{U}n}{2K}$. Here ${\rm u}$ determines the internal phase space structure of the individual dimers, whereas ${\rm v}(\ll {\rm u})$ dictates the energy exchange between neighboring dimers. The classical, mean-field limit of the many body dynamics is attained as $N$ is increased while keeping ${\rm u}$ and ${\rm v}$ fixed, allowing for the replacement of spin operators by $c$-numbers, parametrized as  $s_{x,i}=\sin\theta_i \cos\varphi_i$, $s_{y,i}=\sin\theta_i \sin\varphi_i$, $s_{z,i}=\cos\theta_i$ with $\varphi_i$ and $n\cos\theta_i$ corresponding to the relative phase and population imbalance in site $i$, respectively \cite{vardi,vardi1}. The dimer excitation energy per particle  in site $i$ is given as $E_i=-\frac{K}{2}(s_{x,i}+\frac{u}{2}s^2_{z,i})$ classically or $\langle\frac{K}{2}({\hat s}_{x,i}+\frac{u}{2}{\hat s}^2_{z,i})\rangle$ quantum mechanically, and the classical ground state energy is $E_{min}=-K/2$. The total internal excitation energy $\sum_i (E_i(t)-E_{min})=E$ is conserved to a very good approximation, because the inter-site coupling is weak.

 Following Ref.~\cite{anglin1} we transform the dimer array Hamiltonian to the excitation basis, starting with a Holstein-Primakoff transformation (HPT),
 \begin{eqnarray}
\hat{L}_{x,i}&=&\frac{n}{2}-\hat{A}^{\dagger}_{i}\hat{A}_{i}\equiv \frac{1}{2} (\hat{a}^{\dagger}_{i}\hat{b}_{i}+\hat{b}^{\dagger}_{i}\hat{a}_{i}),\label{hpt1}\\
\hat{L}^{+}_{x,i}&=&\hat{L}_{z,i}-i\hat{L}_{y,i}=\sqrt{n-\hat{A}^{\dagger}_{i}\hat{A}_{i}}\hat{A}_{i}\equiv \frac{1}{2}(\hat{a}^{\dagger}_{i}+\hat{b}^{\dagger}_{i})(\hat{a}_{i}-\hat{b}_{i})
, \label{hpt2} \nonumber\\ \\
\hat{L}^{-}_{x,i}&=&\hat{L}_{z,i}+i\hat{L}_{y,i}=\hat{A}^{\dagger}_{i}\sqrt{n-\hat{A}^{\dagger}_{i}\hat{A}_{i}}\equiv \frac{1}{2}(\hat{a}^{\dagger}_{i}-\hat{b}^{\dagger}_{i})(\hat{a}_{i}+\hat{b}_{i})
,\nonumber 
\label{hpt3}\\
\end{eqnarray}
where $\hat{A}_i$ is the inter-mode atom moving operator with $[\hat{A}_i,\hat{A}^{\dagger}_i]=1$ and $[\hat{A}_i,\hat{n}_i]=0$. The coupling-free Hamiltonian (\ref{ham1}) is thus approximated as, 
\begin{eqnarray}
 H_0&=&\sum_{i}\Big[-K\frac{n}{2}+K\hat{A}^{\dagger}_{i}\hat{A}_{i}+\frac{Un}{4}(\hat{A}_{i}+\hat{A}^{\dagger}_{i})^2
 \nonumber \\&&~~~~~~-\frac{U}{8}\Big\{(\hat{A}_i+\hat{A}^{\dagger}_i),\hat{A}^{\dagger 2}_i\hat{A}_i+\hat{A}^{\dagger}_i\hat{A}^2_i\Big\}\Big]+\textit{O}(Un^{-1})
 .\nonumber \\
 \label{nonint_hpt}
\end{eqnarray}

The local Bogoliubov modes are then obtained by the usual transformation $\hat{c}_i=u_i\hat{A}_i-v_i\hat{A}^{\dagger}_i$, $\hat{c}^{\dagger}_i=u_i\hat{A}^{\dagger}_i-v_i\hat{A}_i$ with $[\hat{c}_i,\hat{c}^{\dagger}_i]=1$ and choosing $u_i$ and $v_i$ so that the quadratic part of Eq. (\ref{nonint_hpt}) is diagonalized.   In the resulting expression, dropping the terms that do not conserve energy of the isolated dimer 
 (i.e., terms not commuting with $\hat{c}^{\dagger}_i\hat{c}_i$), we get the coupling-free Hamiltonian in Bogoliubov basis,
 \begin{eqnarray}
 H_{0}&\approx&\sum_{i}\Big[-\frac{Kn}{2}+\omega_J\hat{c}^{\dagger}_i\hat{c}_i+{U_{J}}\hat{c}^{\dagger}_i\hat{c}^{\dagger}_i\hat{c}_i\hat{c}_i\Big],
\label{nonint_bog}
 \end{eqnarray}
where $\omega_J=\sqrt{K(K+Un)}$ and ${U_{J}}=-\frac{U}{8}\frac{4K+Un}{K+Un}$ are the Josephson oscillation frequency 
 and the effective strength of interaction between the Bogoliubov quasiparticles, respectively. 

Using the same HPT Equations. (\ref{hpt1}), (\ref{hpt2}), and (\ref{hpt3}), the interaction Hamiltonian (\ref{int}) is transformed as,
 \begin{eqnarray}
  H_c&=&\frac{\mathcal{U}}{4}\sum_{<ij>}\Big[(\hat{A}^{\dagger}_i+\hat{A}_i)(\hat{A}^{\dagger}_j+\hat{A}_j)(n-\frac{\hat{A}^{\dagger}_i\hat{A}_i}{2}-\frac{\hat{A}^{\dagger}_j\hat{A}_j}{2})
\nonumber \\ &&~~~~~~~~~~~~~+(\hat{A}^{\dagger}_i+\hat{A}_i)\frac{\hat{A}^{\dagger}_j}{2}+\frac{\hat{A}^{\dagger}_i}{2}(\hat{A}^{\dagger}_j+\hat{A}_j)\Big].
\label{coup_htp}
\end{eqnarray}
Preforming the Bogoliubov transformation and retaining leading, energy conserving terms, the coupling Hamiltonian representation in the excitation basis is,
\begin{eqnarray}
 H_c&\approx&K_{J}\sum_{<ij>}(\hat{c}^{\dagger}_i\hat{c}_j+\hat{c}^{\dagger}_j\hat{c}_i),
\label{coup_bog}
 \end{eqnarray}
where $K_{J}=\frac{\mathcal{U}n}{4\sqrt{1+u}}$ is the effective hopping strength of quasiparticles between neighboring dimers. Thus, by contrast to Ref.~\cite{anglin1} where the site coupling involved the transfer of particles, the coupling here corresponds predominantly to a linear exchange of excitations.

Considering Eq.~(\ref{nonint_bog}) and Eq.~(\ref{coup_bog}), the dimer array Hamiltonian is effectively a BHH for attractively interacting excitations, with $\omega_J$, $K_J$, and $U_J$ functioning as self energy, hopping, and interaction strength, respectively.  This implies the possibility of obtaining discrete two-dimensional excitation solitons,  similar to two-dimensional solitons in photonic lattices \cite{fleischer03} and BECa  \cite{gligoric10}. The characteristic interaction parameter for the self trapping of excitations should thus be, 
\begin{equation}
{\rm u}_J=\frac{|U_J| n_J}{K_J}=\frac{{\rm u}(4+{\rm u})}{4{\rm v}\sqrt{1+{\rm u}}}.\frac{n_J}{n},
\end{equation}
where the approximate number of quasiparticles is deduced by assuming equispaced low energy excitations, to be $n_J\approx\sum_i \frac{n_i(E_i(0)-E_{\rm min})}{\omega_J}$. 

\begin{figure}[t]
\centerline{\includegraphics[width=\linewidth]{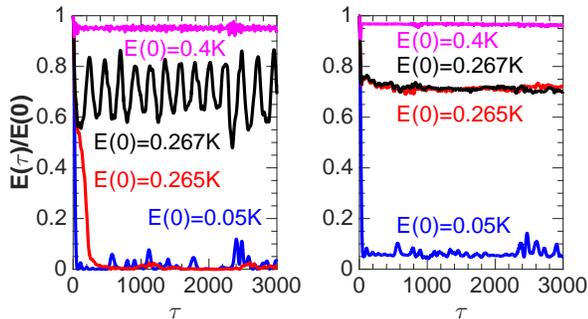}}
\caption{Energy of the initially excited dimer site as a function of the rescaled time $\tau$ for various initial conditions. Mean-field results are plotted in the left panel, whereas full many-body calculations are shown on the right. Simulations were carried out with $n=100$ particles in each site of a $20\times20$ array, with ${\rm u}=1.0$ and ${\rm v}=0.05$. The quantum initial preparations were coherent spin states $|\theta,\varphi\rangle$ corresponding to the same $\{\theta,\varphi\}$ as the classical preparations. Both calculations show self trapping of excitations at high initial excitation energy vs dispersion at low excitation energy.}
\label{fig1}
\end{figure}

 To test for excitation self-trapping, we numerically simulate the dynamics of an extended 2D dimer array.  Both classical (mean-field) and full quantum (many-body) calculations were carried out. Provided that the initial state of the system can be factorized as the product of single-site $n$-particle dimer states, the factorization persists at all times due to the lack of inter-site particle transfer :
\begin{equation}
|\Psi\rangle_{t} =\prod_{i} \left\{\sum_{m_{i}=-\ell}^\ell c_{m_{i}} (t) \left | \ell , m_{i}\right\rangle\right\},
\label{factor}
\end{equation}
where $c_{m_{i}} (t) = \left\langle \ell, m_i| \Psi \right\rangle_t$, and $ \left | \ell , m_{i}\right\rangle$ are the joint eigenstates of ${\hat L}_i^2$ and ${\hat L}_{z,i}$. This greatly reduces the Hilbert space dimension compared with the system considered in \cite{anglin3}, making a numerical calculation possible. The classical simulations were launched with $\{\theta_i,\varphi_i\}_{t=0}$ initial conditions that correspond  to a single excited site, i.e. $E_j(0)-E_{min}= E>0$  and $E_i(0)=E_{min}$ for all $i\neq j$.  The many-body quantum simulations where initiated with a direct product $|\Psi\rangle_{t=0} = \prod_{i} |\theta_{i},\varphi_{i}\rangle$, of spin coherent states $|\theta_{i},\varphi_{i}\rangle$ in each site, corresponding to the same relative phase and population imbalance as in the corresponding classical simulation. 
 
 \begin{figure}[t]
 \centerline{\includegraphics[width=\linewidth]{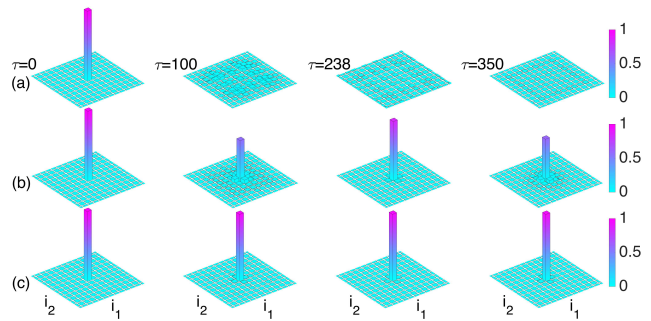}}
\caption{Snapshots of the dimer excitation energy distribution in the array at different times during the classical evolution, for $E/K=0.05$ (a), $0.267$ (b), and $0.4$ (c). When the initial excitation energy is small, energy disperses throughout the array, while larger initial excitation results in breathing and self trapping. Parameters are as in Fig.~\ref{fig1}. For clarity, only the central $11\times11$ sites in the array are shown. Energies are normalized to the initial excited dimer's energy in each case.}
\label{fig2}
\end{figure}

In Fig.~\ref{fig1}, the energy of the excited site, starting at different initial values, is plotted as a function of time. When the initial excitation energy is low, it quickly disperses throughout the array. By contrast, higher excitation energies (a larger initial number of quasiparticles in the excited site) result in the self trapping of energy with periodic exchange of excitations between the initially excited site and its close neighbours.  Finally, for larger excitation energies, we obtain a stationary soliton of excitations. While the classical transition from energy dispersion to self trapping is very sharp, the corresponding quantum crossover is much more moderate. This is due to the finite uncertainty width of the quantum state, which results in the smearing of classical phase space structures (see below). 
 
The evolution of the energy distribution in the array during the mean-field simulation for three typical initial conditions, is plotted in Fig.~\ref{fig2}. For small initial energy in the excited site, energy disperses rapidly through the array. However, beyond a critical value, the excitation energy remains self trapped with breathing of the energy distribution. As the initial excitation energy is further increased, the breathing encompasses a narrower environment  of the initially excited site, until a stationary 2D energy soliton is obtained.
 
 \begin{figure}[t]
 \centerline{\includegraphics[width=\linewidth]{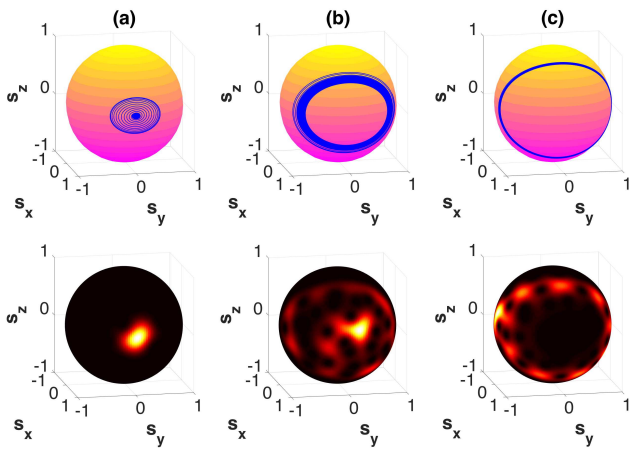}}
\caption{Classical phase-space trajectories for the initially excited dimer (top) and the quantum Hussimi phase space distribution at $\tau=400$ (bottom) for the same parameters as in Fig.~\ref{fig2}.}
\label{fig3}
\end{figure}                                              

The classical phase space trajectories ${\bf s}_i(t)$ in the initially excited site, are presented in the upper panels of Fig. \ref{fig3}.  Conservation of ${\bf s}^2$ implies classically that the norm $s$ is conserved and the mean-field trajectories are restricted to the surface of the unit sphere, in accordance with the classical minimal-Gaussian ansatz of a pure state throughout the evolution \cite{vardi,vardi1}. In agreement with the previous figures, a low energy excitation decays to the classical ground state (${\bf s}=(1,0,0)$) whereas higher excitations result in beating between two limiting excited trajectories and eventually full self trapping on the initial energy contour. 
                                                                                     
The corresponding Husimi distribution $P(\theta,\varphi)=\left | \langle \theta,\varphi | \Psi\rangle\right |^2$ for the many-body quantum calculation at  $\tau=400$  is shown in the lower panels of Fig. \ref{fig3}. Unlike the classical case, conservation of $\langle{\bf s}^2\rangle$ does not imply constant $|\langle {\bf s}\rangle|$ due to the finite quantum variance of $\hat{\bf s}$ and the full quantum state looses its one-particle purity during the evolution \cite{vardi,imamoglu,wright}. If the relaxation to the coherent ground state is fast, the initial excited coherent state remains close to a minimal-uncertainty Gaussian and retains its one-particle purity. By contrast, when the excitation energy is self trapped, the quantum phase-space distribution spreads all over the corresponding classical trajectory, and one-particle coherence is lost. This smearing suppresses the classical beating, as the distribution spreads to the entire region between the limiting energy contours rather than coherently propagating between them. For intermediate initial excitation energy, part of the initial coherent quantum distribution still has subcritical energy and hence decays, whereas the remaining part is self trapped, resulting in a bi-modal final distribution (hence the gradual quantum transition from relaxation to self trapping).  The blurring of the classical features decreases with $n$, as the size of the initial phase-space distribution becomes smaller and the mean-field limit is approached.

 \begin{figure}[t]
 \centerline{\includegraphics[width=0.8\linewidth]{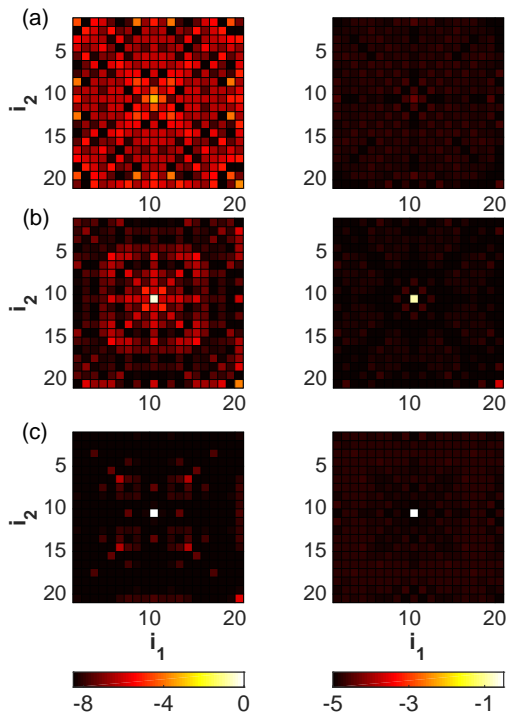}}
\caption{Energy distribution $E_i(\tau)$ and one-particle entropy distribution ${\cal S}_i(\tau)$ at the end of the quantum simulation ($\tau=3000$), for the same initial excitation energies and parameters as in Fig.~\ref{fig2}. Note that the color map is scaled logarithmically.}
\label{fig4}
\end{figure}

In Fig.~\ref{fig4} we plot the distribution of energy (left column) and of the entanglement entropy (right column),
\begin{equation}
{\cal S}_{i}={\rm Tr} \left(\rho_{i}^{(sp)}\ln \rho_{i}^{(sp)}\right)=-\frac{1}{2}\ln\left(\frac{(1+s_{i})^{1+s_{i}}(1-s_{i})^{1-s_{i}}}{4}\right)~,
\end{equation}
at the initially populated site, at the end of the propagation for the three representative excitation energies of the previous figures. Here, $\rho_{i}^{(sp)}$ is the reduced one particle density matrix of the $i$-th dimer. It is clear that due to loss of one particle coherence which accompanies self-trapping, the new localized structures are not only energy solitons but also entanglement-entropy solitons. In addition to being fundamentally important  from a thermodynamical perspective, this feature may be useful for quantum information processing applications requiring quantum entanglement as a resource. 

To conclude, using a two-dimensional array of Bose-Hubbard dimers with a weak nearest-neighbor interaction, we have demonstrated the macroscopic self-trapping of energy, with Josephson quasiparticles replacing the atoms of the standard schemes. The resulting quasiparticle solitons feature localization of both energy and entropy, while the particle density is uniform. The existence of robust many-body states that do not dissipate in experimentally realizable settings, would be helpful in designing experiments related to lossless quantum-mechanical architectures and hysteresis-based quantum memories \cite{trenkwalder,eckel}.

\end{document}